\documentclass[aps,pre,preprint,groupedaddress,superscriptaddress,showpacs,a4paper,floatfix]{revtex4-1}

\usepackage{amsmath,amssymb,graphicx,graphics,bm}
\usepackage{hyperref}
\usepackage{graphics,graphicx}
\usepackage{amsmath,amssymb}

\begin{document}

\title{On the ergodicity breaking in well-behaved Generalized Langevin Equations}

\author{Giuseppe Procopio,  Chiara Pezzotti and Massimiliano Giona}
\email[corresponding author:]{massimiliano.giona@uniroma1.it}
\affiliation{Dipartimento di Ingegneria Chimica, Materiali, Ambiente La Sapienza Universit\`a di Roma\\ Via Eudossiana 18, 00184 Roma, Italy}

\date{\today}

\begin{abstract}
The phenomenon of ergodicity breaking of stochastic dynamics governed by Generalized Langevin Equations (GLE)
in the presence of well-behaved  exponentially decaying dissipative memory kernels,
recently investigated by many authors 
(Phys.  Rev. E {\bf 83} 062102 2011;  Phys. Rev. E {\bf 98} 062140 2018;
Eur. Phys. J. B {\bf 93} 184 2020),
 finds,
in the dynamic theory of GLE,
its simple and natural explanation, related to the concept of dissipative stability.
It is shown that the occurrence of ergodicity breakdown for well-behaved dissipative kernels falls, in general, 
ouside the region of
stochastic realizability, and therefore it cannot be  observed in physical systems.
\end{abstract}

\maketitle

\noindent
\section{Introduction}  
Since the work by Kubo \cite{kubo1,kubo2},  the use of Generalized Langevin Equations (GLE)
to describe physical systems driven by thermal fluctuations, their equilibrium properties and their
response to external perturbations, has become a standard  procedure that applies
in the linear regime from Brownian dynamics to
dielectric response, from electrochemical processes to magnetization phenomena.
If $A(t)$ is the physical observable, the GLE describing its linear dynamics at thermal equilibrium 
takes the form
\begin{equation}
 \frac{d A(t)}{ d t}= - \int_0^t M(t-\tau) \, A(\tau) \, d \tau + R(t)
\label{eqm_1}
\end{equation}
where  $M(t)$ is a memory  kernel accounting for dissipation,
and $R(t)$ is the stochastic forcing associated with the thermal fluctuations.
In order to  focus the idea on a concrete physical problem, consider the case where $A(t)= m \, v(t)$ is the momentum
of a (spherical) particle of mass $m$ in a fluid medium, so that eq. (\ref{eqm_1}) can be rewritten  in terms of the particle
velocity $v(t)$ as
\begin{equation}
m \, \frac{d v(t)}{d t}=- \int_0^t h(t-\tau) \, v(\tau) \, d \tau + R(t)
\label{eqm_2}
\end{equation}
where $h(t)=m \, M(t)$,
and eq. (\ref{eqm_2}) is coupled with  the kinematic equation
\begin{equation}
\frac{d x(t)}{d t}= v(t)
\label{eqm_3}
\end{equation}
As regards $R(t)$, the classical assumption in   GLE theory is that the thermal fluctuations are independent of the
velocity, and this  can be expressed by the Langevin condition \cite{langevin}
\begin{equation}
\langle R(t) \, v(0) \, \rangle_{\rm eq}=0 \, , \qquad t \geq 0
\label{eqm_4}
\end{equation}
where $\langle \cdot \rangle_{\rm eq}$ indicates the expected value with respect to the
probability measure of the thermal fluctuations at equilibrium (at constant temperature $T$).

The kernel $h(t)$  accounts for the dissipative nature of the interactions that,  combined with the
fluctuating contribution $R(t)$, determines the relaxation towards equilibrium.
In the case of  the equation of motion for a Brownian particle eq. (\ref{eqm_2}), the kernel $h(t)$
represents the dynamic friction factor \cite{procopio2}, proportional to the
dynamic viscosity of the medium \cite{rheol1,rheol2}. For a Newtonian fluid in the Stokes regime
$h(t)$ is impulsive and this corresponds to the classical Einstein-Langevin description
of Brownian motion. For more complex fluids, viscoelastic effects becomes important
at the time-scales of interest (both in theory and in the applications) \cite{franosch}
and $h(t)$ becomes a continuous function of time.

The basic condition in the classical Kubo theory to be imposed on $h(t)$ is that the long-term
friction coefficient $\eta_\infty$,
\begin{equation}
\eta_\infty = \int_0^\infty h(t) \, d t  \geq 0
\label{eqm_5}
\end{equation}
is positive \cite{kubo1,kubo2}. More precisely,  three cases may occur: (i) if $\eta_\infty >0$ and bounded,  then
particle dynamics is in the long-term diffusive, i.e. $\langle x^2(t) \rangle \simeq 2 \, D \, t$,
where the diffusivity $D$ is related to $\eta_\infty$ by the global fluctuation-dissipation relation
\begin{equation}
D \, \eta_\infty = k_B \, T
\label{eqm_6}
\end{equation}
where $k_B$ is the Boltzmann constant; (ii) $\eta_\infty=\infty$, and in this case anomalous subdiffusive
behavior  may occur; (iii) $\eta_\infty=0$, corresponding to the absence of any dissipation, and this
gives rise either to anomalous superdiffusive scaling or the breaking or ergodicity \cite{goychuk1,goychuk2,goychuk3}.

This picture has been questioned by a series of interesting contributions \cite{bao1,bao2,bao3,plyukhin1,plyukhin2}
that showed that, even if $\eta_\infty$ is positive and finite and  the kernel $h(t)$ possesses
a regular exponentially decaying behavior with $t$,  (i.e., $|h(t)| \leq C \, e^{-k\, t}$, for some $C>0$ and $k>0$),
ergodicity breaking may occur, associated with a non-vanishing long-term scaling of
the velocity autocorrelation function $C_{vv}(t)= \langle v(t) \, v(0) \rangle_{\rm eq}$.
The latter property implies that there exists a monotonically increasing and diverging sequence
of time instants $\{ t_n \}_{n=1}^\infty$, $t_n < t_{n+1}$, $\lim_{n \rightarrow \infty} t_n =\infty$,
and a  constant $K>0$ such that $|C_{vv}(t_n)| \geq K$.
We use the diction of ``well-behaved'' dissipative kernels for  $h(t)$'s  possessing the following properties: (i) $\eta_\infty>0$;
(ii) $h(t)$ is an exponentially decaying function of $t$ for large times; (iii) $h(t) \geq 0$.
Condition (iii) usually emerges as a physical empirical constraint (i.e. dictated by the
phenomenological experience) on the viscoelastic 
response of complex fluids. The model considered in \cite{bao3} 
fulfils conditions (i) and (ii) but not (iii). Nevertheless it is  also considered  below for its mathematical interest,
as a non-hydrodynamic model of relaxation.

In point of fact, the results discussed in \cite{bao1,bao2,bao3,plyukhin1,plyukhin2}
 on the ergodicity breaking in well-behaved dissipative
systems can be viewed as a particular case of the dynamic theory developed in \cite{gpp} for GLE, in which
it is shows that eq. (\ref{eqm_5}) is by no mean a sufficient condition to establish  the relaxation towards 
a stable equilibrium behavior and the validity of
the global Stokes-Einstein relation  eq. (\ref{eqm_6}), as intrinsic  instabilities in the
internal memory dynamics governed by the kernel $h(t)$ may occur, leading to a diverging
behavior in time of the velocity autocorrelation function.
More precisely, the ergodicity breaking observed in  \cite{bao1,bao2,bao3,plyukhin1,plyukhin2} corresponds
to operating conditions lying at the boundary of the region of (dissipative) stability as defined
in \cite{gpp}. Following the analysis developed in \cite{gpp}, in order to observe
these phenomena from the dynamic evolution of physical systems  described via  GLE, the
further condition of stochastic realizability should be met.
As  will be  thoroughly addressed below, the phenomena of ergodicity breaking  reported in
\cite{bao1,bao2,bao3,plyukhin1,plyukhin2} cannot occur in general  due to the lack of stochastic
realizability \cite{gpp}.
The aim of this brief report is to address in detail these issues in full clarity, by analyzing two
benchmarking examples: kernels $h(t)$ possessing two real-valued relaxation
exponents, and the model addressed by Plyukhin in \cite{bao3}.\\

\section{Dynamic GLE theory} 
To begin with, let us review the main results developed in \cite{gpp} that
are functional to the present analysis. Without loss of generality, asume that $h(t)$ can be expressed as
a series of exponentially decaying functions of times (modes) plus an impulsive contribution
\begin{equation}
h(t)= \sum_{k=1}^N h_k \, e^{-\lambda_k \, t} +  H_0 \, \delta (t)
\label{eqm_7}
\end{equation}
where $\lambda_k \in {\mathbb C}$, $k=1,\dots,N$ are complex-valued exponents, and consequently $h_k \in {\mathbb C}$,
such  that the sum entering eq. (\ref{eqm_8}) is real-valued for any $t\geq 0$, and $H_0>0$.  Moreover, let us
assume that $\mbox{Re}[\lambda_k]>0$, $k=1,\dots,N$, so that $h(t)$ decays exponentially with $t$, and that
\begin{equation}
\eta_\infty = H_0 + \sum_{k=1}^N \frac{h_k}{\lambda_k} >0
\label{eqm_8}
\end{equation}
and bounded, where obviously the sum in eq. (\ref{eqm_8}) is real-valued.

Eq. (\ref{eqm_7}) essentially corresponds to the existence of a Markovian embedding for the GLE.
It has been argumented in \cite{gpp} that this representation, considering the limit for $N \rightarrow \infty$, constitutes the most
general structure for the linear response of a physical system that is local in time.
The only exception to this claim occurs  in some particular cases where the particle interacts with a field,
e.g. the  hydrodynamic velocity field of a continuous liquid phase, determining in the Newtonian
case the occurrence of a fluid-inertial memory  contribution to $h(t)$ proportional to $1/\sqrt{t}$ (the Basset force) \cite{landau}.
But even this case reduces to eq. (\ref{eqm_7}) (with a countable set of modes) if
the finite propagation velocity of the shear stresses is accounted for \cite{procopio2}.
In any case, a further discussion on this important conceptual point is completely immaterial in the present
analysis as all the examples considered in \cite{bao1,bao2,bao3,plyukhin1,plyukhin2}
refer to the modal representaton eq. (\ref{eqm_7}) with $N$ finite and small.

Given eq. (\ref{eqm_7}), the GLE can be represented as
\begin{eqnarray}
m \, \frac{d v(t)}{d t} & = & - \sum_{k=1}^N \widetilde{h}_k \, z_k(t) - H_0 \, v(t) + \sqrt{2} \, \beta_{0,0}
\, \xi_0 (t) + \sqrt{2} \, \sum_{k=1}^N \beta_{0,k} \, \xi_k(t) \nonumber \\
\frac{d z_k(t)}{d t} & = & - \lambda_k \, z_k(t) + b_k \, v(t) + \sqrt{2} \, \beta_{k,0} \, \xi_0(t) +
\sqrt{2} \, \sum_{h=1}^N \beta_{k,h} \, \xi_k(t)
\label{eqm_9}
\end{eqnarray}
where $\xi_0(t)$, $\xi_k(t)$, $k=1,\dots,N$ are independent white-noise processes \cite{vankampen},
$\langle \xi_h(t) \, \xi_k(t^\prime) \rangle = \delta_{h,k} \, \delta(t-t^\prime)$, $h,k=0,\dots,N$, $t, \, t^\prime \in {\mathbb R}$,
e.g. distributional
derivatives of independent Wiener processes (corresponding to the classical approach adopted in statistical
physics and followed also in the present work), and
\begin{equation}
b_k \, \widetilde{h}_k = h_k \, , \qquad k=1,\dots,N
\label{eqm_10}
\end{equation}
The coefficients $\beta_{h,k}$, $h,k=0,\dots,N$, modulating the intensity of the stochastic fluctuations,  are to be determined from the Langevin condition eq. (\ref{eqm_4}),
indicating that the thermal fluctuations at any time instant $t$ are independent of the previous history of particle
velocity. This leads   automatically to the equations for the velocity autocorrelation function $C_{vv}(t)$
\begin{eqnarray}
m \, \frac{d C_{vv}(t)}{ dt}  & =  & - \sum_{k=1}^N \widetilde{h}_k \, C_{z_k v}(t) \nonumber \\
\frac{d C_{z_k v}(t)}{ dt}  & =  & - \lambda_k \, C_{z_k v}(t) + b_k \, C_{vv}(t) \, , \qquad k=1,\dots, N
\label{eqm_11}
\end{eqnarray}
where $C_{z_k v}(t)= \langle z_k(t) \, v(0) \rangle_{\rm eq}$, $k=1,\dots,N$, equipped with the initial
conditions
\begin{equation}
C_{vv}(0)= \langle v^2 \rangle_{\rm eq}= \frac{k_B \, T}{m} \, , \qquad C_{z_k v}(0)=0 \, \quad k=1,\dots,N
\label{eqm_12}
\end{equation}
Eqs. (\ref{eqm_11})-(\ref{eqm_12}) are usually referred to as the fluctuation-dissipation relation (theorem, in the Kubo
description \cite{kubo1,kubo2}) of the first kind.

The structure of the stochastic perturbations entering eq. (\ref{eqm_9}), and its white-noise nature
represents the most general setting consistent with the Langevin conditions eq. (\ref{eqm_4}).
Henceforth, let us consider the nondimensional formulation of the equations of motion, rescaling the velocity
to its equilibrium intensity, and time with respect to a characteristic dissipation time. 
In practice, and without loss of generality, this means that we can set $m=1$, $\langle v^2 \rangle_{\rm eq}=1$,
while the remaining coefficients $\widetilde{h}_k$, $H_0$, $\lambda_k$, $b_k$ are dimensionless.
Setting ${\bf y}=(v,z_1,\dots,z_N)$,  $\boldsymbol{\xi}=(\xi_0,\xi_1,\dots,\xi_N)$ eq. (\ref{eqm_9}) can be
compactly expressed as
\begin{equation}
\frac{d {\bf y}(t)}{d t}= {\bf A} \, {\bf y}(t) + \sqrt{2} \, \boldsymbol{\beta} \, \boldsymbol{\xi}(t) 
\label{eqm_13}
\end{equation}
where the coefficient matrix ${\bf A}$ governing the internal mode dynamics is given by
\begin{equation}
{\bf A}=
\left (
\begin{array}{ccccc}
-H_0 & - \widetilde{h}_1 & \dots & \dots & -\widetilde{h}_N \\
b_1 & -\lambda_1 & 0 & \dots & 0 \\
b_2 & 0 & -\lambda_2 & \dots & 0 \\
\dots & \dots & \dots & \dots & \dots \\
b_N & 0 & 0 & \dots & -\lambda_N 
\end{array}
\right )
\label{eqm_14}
\end{equation}
$\boldsymbol{\beta}=(\beta_{h,k} )_{h,k=0}^N$,  and  $\boldsymbol{\beta} \, \boldsymbol{\xi}(t)$
corresponds to the row-by-column matrix multiplication of vector  $\boldsymbol{\xi}(t)$
by  the matrix $\boldsymbol{\beta}$.

In order to enforce fluctuation-dissipation relations, two basic properties should be
fulfilled \cite{gpp}, namely
\begin{itemize}
\item {\em dissipative stability}, corresponding to the fact that all the eigenvalues $\nu_k$, $k=0,\dots, N$ of
the coefficient matrix ${\bf A}$ possess  negative real part;
\item {\em stochastic realizability}, corresponding to the fact that there exists a matrix $\boldsymbol{\beta}$
of stochastic intensities such as, given  the stochastic process eq. (\ref{eqm_10}), the fluctuation-dissipation
relation of the first kind, i.e., eqs.  (\ref{eqm_11})-(\ref{eqm_12}) for $C_{vv}(t)$, is fulfilled.
\end{itemize}
If the system is dissipatively unstable, i.e., if there exists at least one eigenvalue of ${\bf A}$, say $\nu_{k^*}$,
with $\mbox{Re}[\nu_k^*]>0$, no equilibrium conditions can be set. As a consequence, no diffusive dynamics exists
and the Stokes-Einstein relations  cannot be applied. But even if a GLE is dissipatively stable,
this does not necessarily implies that the thermal fluctuations could be expressed in the form
of a stochastic process $R(t)$ entering eq. (\ref{eqm_2}) and such that it fulfils the Langevin condition
eq. (\ref{eqm_4}). In order to ensure  it, the requirement of stochastic realizability
should be further enforced.

In the light of the phenomenology envisaged by the ergodicity breaking addressed in
 \cite{bao1,bao2,bao3,plyukhin1,plyukhin2}, the dynamic theory of GLE simply indicates that
this occurs at the boundary separating the region of parameters for which the GLE is dissipatively stable
from the instability region. 

In the remainder we consider two prototypical examples, showing  in general that the condition of stochastic
realizability is more stringent than dissipative stability, and consequently, in most of the situations,
a stochastic GLE with well-behaved kernel providing ergodicity breaking cannot be defined within the realm of
the actual theory of  GLE, in the meaning that there is no stochastic process $R(t)$ fulfilling the Langevin condition eq. (\ref{eqm_4}), such that the 
resulting
GLE eq. (\ref{eqm_2}) satisfy the fluctuation-dissipation relation of the first kind.\\

\section{Two real-mode  dynamics} 

To begin with, consider the case of two real modes, i.e.,
\begin{equation}
h(t)=h_1 \, e^{-\lambda_1 \, t} + h_2 \, e^{-\lambda_2 \, t}
\label{eqm_15}
\end{equation}
setting $h_1=1$, $\lambda_1,\lambda_2 >0$, $\lambda_1 <\lambda_2$.
The case of real modes occurs in the study of particle motion in linear viscoelastic fluids, where the
values of the coefficients/exponents, $h_k$ and $\lambda_k$, respectively,
 can be determined from the analysis
of rheological experiments \cite{rheol1,rheol2}. In the present context, the case of major interest is where $h_2=-\alpha<0$,
as  the system is both dissipatively stable and stochastically
realizable for any non-negative values of $h_k$. From eq. (\ref{eqm_15}), setting without loss of generality,
 all the $b_k's$ equal to 1 in eq. (\ref{eqm_9}), the expression of the coefficient matrix ${\bf A}$ becomes
\begin{equation}
{\bf A}=
\left (
\begin{array}{ccc}
0 & - h_1 & \alpha \\
1 & - \lambda_1 & 0 \\
1 & 0 & -\lambda_2
\end{array}
\right )
\label{eqm_16}
\end{equation}
and $\alpha$ is used as a parameter controlling the  qualitative properties of the GLE.
Following the  dynamical theory of GLE \cite{gpp}, three different characteristic values
of $\alpha$ can be defined. The first one is the threshold $\alpha^*$, corresponding to
the situation where $\eta_\infty |_{\alpha=\alpha^*}=0$, i.e.,
\begin{equation}
\alpha^* = \frac{\lambda_2 \, h_1}{\lambda_1}
\label{eqm_17}
\end{equation}
For $\alpha< \alpha^*$ $\eta_\infty >0$, while the global dissipative properties
are lost for $\alpha> \alpha^*$ and in this region the fluid acts as an actively destabilizing environment
for particle dynamics.
The second critical value  $\alpha_{\rm ds}$ determines the boundary of the region of  dissipative stability,
as for $\alpha> \alpha_{\rm ds}$ the system is  unstable.
In the present case, $\alpha_{\rm ds}$ can be calculated in an easy way, observing that the
matrix ${\bf A}$ possesses one real eigenvalue and  a couple of complex conjugate eigenvalues, and
that instability originates from the complex-conjugate branch (this
follows from the direct spectral analysis of the system). Letting $\mbox{Tr}$, $M_2$ and $\mbox{Det}$ be the
three algebraic invariants of ${\bf A}$,
\begin{equation}
\mbox{Tr}=-(\lambda_1+\lambda_2) \, , \quad M_2=\lambda_1 \, \lambda_2 + h_1+ h_2 \, ,
\quad
\mbox{Det}= - (h_1 \, \lambda_2 + h_2 \, \lambda_1)
\label{eqm_18}
\end{equation}
the critical value $\alpha_{\rm d s}$ of $\alpha$ (cfr. $h_2=-\alpha$) occurs when $\mbox{Det}= \mbox{Tr} \, M_2$,
corresponding to the condition where the complex eigenvalue branch admits vanishing real part.
This leads to the expression for $\alpha_{\rm ds}$,
\begin{equation}
\alpha_{\rm ds}= - \frac{1}{\lambda_2} \, \left [ h_1 \, \lambda_2 - (\lambda_1+\lambda_2) ( \lambda_1 \, \lambda_2 +h_1) \right ]
\label{eqm_18bis}
\end{equation}
Finally, the last critical value corresponds to $\alpha_{\rm sr}$,  determining the upper value of $\alpha$  at which
the GLE is stochastically realizable.
By considering the general structure of the stochastic perturbation in eq.  (\ref{eqm_9}) (with $H_0=0$  and
$m=1$),  the most general stochastic realization of the system takes the form
\begin{eqnarray}
\frac{d v(t)}{d t} & = & - h_1 \, z_1(t) +\alpha \, z_2(t) \nonumber \\
\frac{d z_1(t)}{d t} & = & - \lambda_1 \, z_1(t) + v(t) + \sqrt{2} \left [ \beta_{1,1} \, \xi_1(t) + \beta_{1,2} \, \xi_2(t)
\right ]
\label{eqm_19} \\
\frac{d z_2(t)}{d t} & = & - \lambda_2 \, z_2(t) + v(t) + \sqrt{2} \left [ \beta_{2,1} \, \xi_1(t) + \beta_{2,2} \, \xi_2(t)
\right ] \nonumber 
\end{eqnarray}
Observe that  no direct coupling between $v(t)$ and a stochastic forcing may occur, since in present case no
impulsive friction is present, and a direct stochastic forcing acting on $v(t)$ would determine the unbounded divergence of the
velocity variance in time.
The Fokker-Planck equation for the probability density $p(v,z_1,z_2,t)$ associated with eq. (\ref{eqm_19})  reads
\begin{eqnarray}
\frac{\partial p}{\partial t} & = & \left (h_1 \, z_1 - \alpha \, z_2 \right ) \, \frac{\partial p}{\partial v}
- \frac{\partial \left [(-\lambda_1 \, z_1+ v) \, p \right ]}{\partial  z_1} 
- \frac{\partial \left [(-\lambda_2 \, z_2+ v) \, p \right ]}{\partial  z_2 } \nonumber \\
& + & S_{1,1} \, \frac{\partial^2 p}{\partial z_1^2} + 2 \, S_{1,2} \, \frac{\partial^2 p}{\partial z_1 \partial z_2}
+ S_{2,2} \, \frac{\partial^2 p}{\partial z_2^2}
\label{eqm_20}
\end{eqnarray}
where ${\bf S}=(S_{i,j} )_{i,j=1}^2$ is given by ${\bf S}=\boldsymbol{\beta} \, \boldsymbol{\beta}^T$,
$\boldsymbol{\beta}=(\beta_{i,j} )_{i,j=1}^2$
and ``$T$''  indicates  the transpose. The GLE  is stochastically realizable if there exists a $2 \times 2$ positive
 definite matrix ${\bf S}$
 entering eq. (\ref{eqm_20}), such that the stationary (equilibrium) second-order moments satisfy the
relations $\langle v^2 \rangle_{\rm eq}=1$, $\langle z_k \, v \rangle_{\rm eq}=0$, $k=1,2$, that correspond to the
fulfillment of the fluctuation-dissipation relation of the first kind. Second-order moments are considered as the first-order
moment identically vanish at equilibrium.
The details of this calculation can be found in \cite{gpp} and therefore are not repeated here. The condition
of positive definiteness  of ${\bf S}$ implies that there exists values of $\zeta=S_{1,2}$ such that
\begin{equation}
\phi(\zeta)= -1 + (\gamma_2-\gamma_1) \, \zeta - \left ( \gamma_3- \gamma_1 \, \gamma_2 \right ) \, \zeta \geq 0
\label{eqm_21}
\end{equation}
where
\begin{equation}
\gamma_1 = - \frac{2 \, h_2}{\lambda_1+\lambda_2} \, , \quad \gamma_2= \frac{2 \, h_1}{\lambda_1+\lambda_2} \, ,
\quad \gamma_3= - \frac{h_1 \, h_2}{\lambda_1 \, \lambda_2}
\label{eqm_22}
\end{equation}
Therefore, the critical value of $\alpha$, as regards stochastic realizability, corresponds to the
situation when the local (and absolute) maximum  $\phi^*$ of the function $\phi(\zeta)$ vanishes.
As
\begin{equation}
\phi^* = \frac{(\gamma_1+\gamma_2)^2 - 4 \, \gamma_3}{4 \, (\gamma_3  - \gamma_1 \, \gamma_2)}
\label{eqm_23}
\end{equation}
this implies $(\gamma_1+\gamma_2)^2-4 \, \gamma_3=0$, that solved with respect to $\alpha=\alpha_{\rm sr}$, provides
\begin{equation}
\alpha_{\rm sr}= - \left ( \delta - \sqrt{\delta^2 - h_1^2} \right ) \, , \qquad \delta= h_1- \frac{h_1 \, (\lambda_1+\lambda_2)^2}{2 \, \lambda_1 \, \lambda_2}
\label{eqm_24}
\end{equation}

\begin{figure}
\includegraphics[width=10cm]{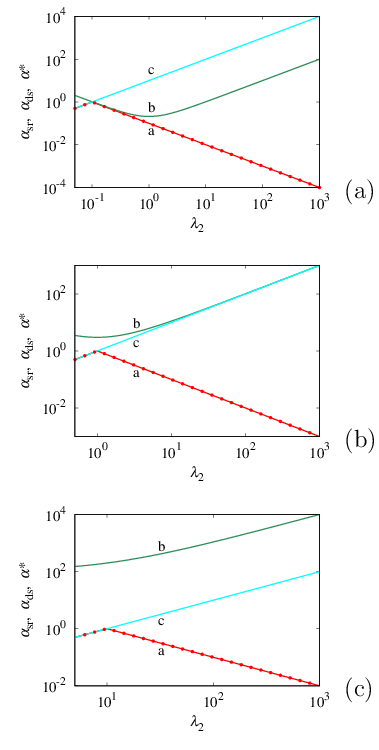}
\caption{$\alpha_{\rm sr}$ (lines a and symbols $\bullet$),  $\alpha_{\rm ds}$ (lines b) and $\alpha^*$ (lines c)
vs $\lambda_2$,  at three
different values of $\lambda_1$ with $h_1=1$. Panel (a) refers to $\lambda_1=0.1$, panel (b) to $\lambda_1=1$, panel (c)
to $\lambda_1=10$.}
 \label{Fig1}
\end{figure}

\begin{figure}
\includegraphics[width=10cm]{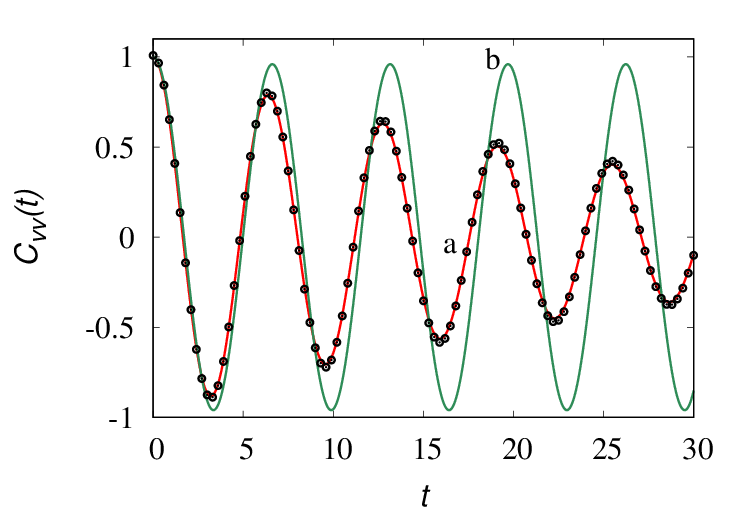}
\caption{Velocity autocorrelation function $C_{vv}(t)$ vs $t$ for the two-mode system discussed in the main
text at $\lambda_1=0.1$, $\lambda_2=1.5$, $\alpha=0.066 < \alpha_{\rm sr}$. Symbols ($\circ$) represent the results of stochastic
simulations, line (a) corresponds to the solution  of eqs. (\ref{eqm_11})-(\ref{eqm_12}) in the present case. Curve (b)
corresponds to the autocorrelation function at $\alpha_{\rm ds}$.} 
\label{Fig2}
\end{figure}

\begin{figure}
\includegraphics[width=10cm]{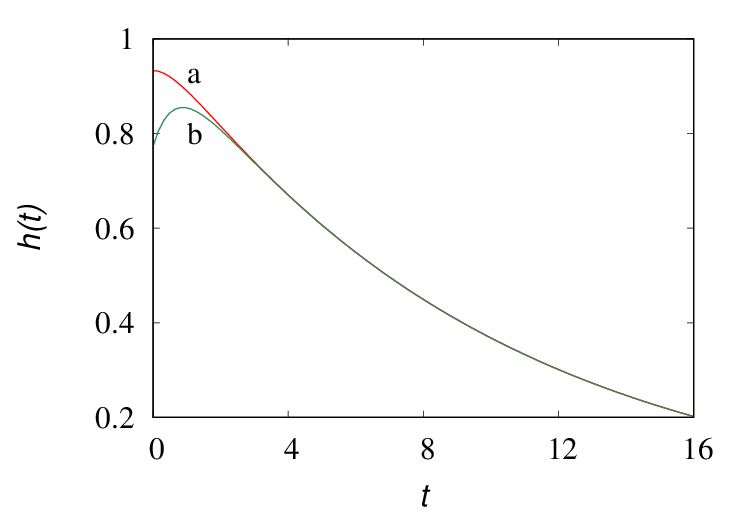}
\caption{Memory kernels $h(t)$ vs $t$ for $\lambda_1=0.1$, $\lambda_2=1.5$ at the critical values.
Line (a) refers to $\alpha=\alpha_{\rm sr}$, line (b) to $\alpha=\alpha_{\rm ds}$.}
\label{Fig3}
\end{figure}

Figure \ref{Fig1} depicts the behavior of $\alpha_{\rm sr}$, $\alpha_{\rm ds}$ and $\alpha^*$ vs $\lambda_2$ at three
different values of $\lambda_1$ (with $h_1=1$). 
As can be observed, in all the situations
\begin{equation}
\alpha_{\rm sr} \leq \mbox{min} \{ \alpha_{\rm ds}, \alpha^* \}
\label{eqm_25}
\end{equation}
and thus,  in all the cases, $\alpha_{\rm sr} < \alpha_{\rm ds}$. This observation is relevant in the
present analysis of the ergodicity breaking as it indicates that prior to observing a non ergodic behavior (corresponding
to the conditions lying on the curves defining $\alpha_{\rm ds}$), the GLE becomes stochastically non realizable.
Consequently, the ergodicity breaking cannot be observed from  the stochastic dynamics of this system.
As an example consider the case $\lambda_1=0.1$, $\lambda_2=1.5$. In this case, $\alpha_{\rm sr}=2/30=0.0666..$, and $\alpha_{\rm ds}=34/15\simeq 0.2266..$. Figure \ref{Fig2} depicts the velocity autocorrelation function $C_{vv}(t)$ for  a value of $\alpha=0.066$ below but
close to $\alpha_{\rm sr}$. Symbols correspond to the stochastic simulations of the GLE, obtained by selecting
a stochastic realization amongst all the possible equivalent ones  at this value of $\alpha$
close to $\alpha_{\rm sr}$,
 considering a symmetric
$\boldsymbol{\beta}$, and  solving the matrix equation ${\bf S}=\boldsymbol{\beta}^2$.
Curve (b) in figure \ref{Fig2} represents the velocity autocorrelation function at $\alpha=\alpha_{\rm ds}$ i.e.
at the condition of ergodicity breaking, when  $C_{vv}(t)$ asymptotically oscillates without neither decaying to
zero or diverging to infinity.
It is worth observing that the phenomena of dynamic instability and stochastic irrealizability observed in
this two-mode system correspond to extremely regular and well-behaved memory kernels, as depicted in
figure \ref{Fig3}.
In the region between $\alpha_{\rm sr}$ and $\alpha_{\rm ds}$ the kernels attains positive values,
and of course decay exponentially to zero for $t \rightarrow \infty$.
Therefore, as  addressed in \cite{gpp}, the dynamic instabilities occurring for this class of  GLE involve a much finer
interaction mechanism amongst the internal modes accounting for the memory dynamics.

\section{The Plyukhin dynamics} 

\begin{figure}
\includegraphics[width=10cm]{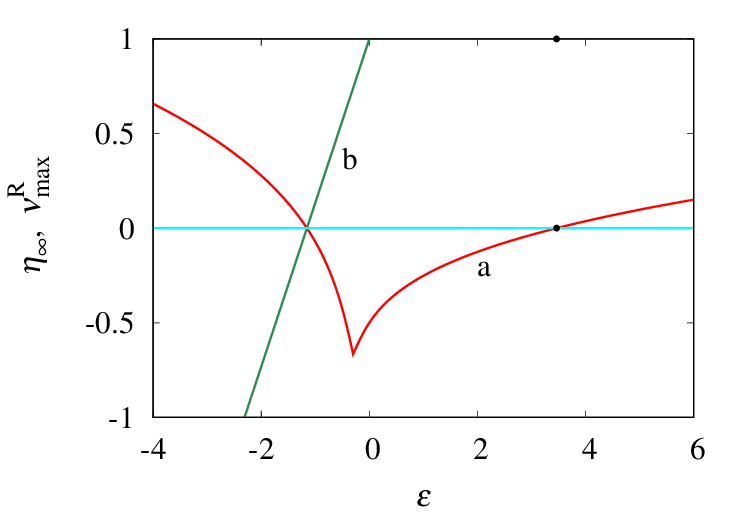}
\caption{$\nu_{\rm max}^R$ (line a) and $\eta_\infty$ (line b) vs $\varepsilon$ for the Plyukhin model
discussed in the main text. The symbol ($\bullet$), corresponding to the
intesection of line (a) with the $x$-axis, marks  the
critical value $\varepsilon^*$.}
\label{Fig4}
\end{figure}

\begin{figure}
\includegraphics[width=10cm]{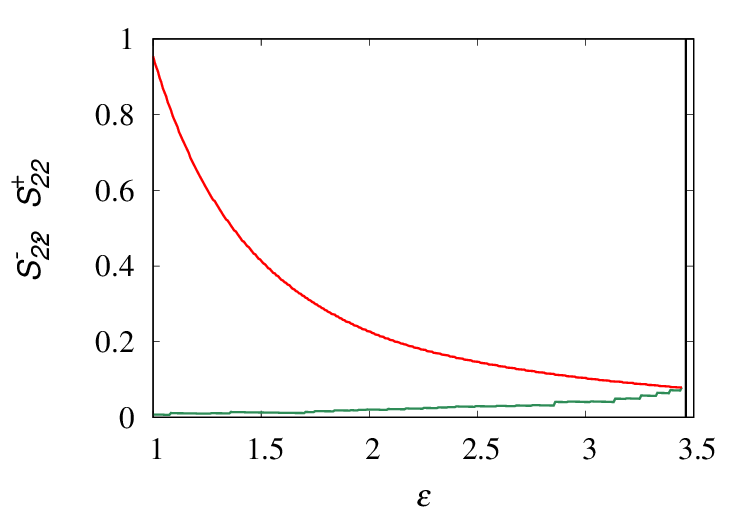}
\caption{$S_{2,2}^-$ (lower curve) and $S_{2,2}^+$ (upper curve) vs $\varepsilon$ for the Plyukhin system discussed in the main text.
The vertical line marks the value of $\varepsilon^*$.}
\label{Fig5}
\end{figure}

\begin{figure}
\includegraphics[width=10cm]{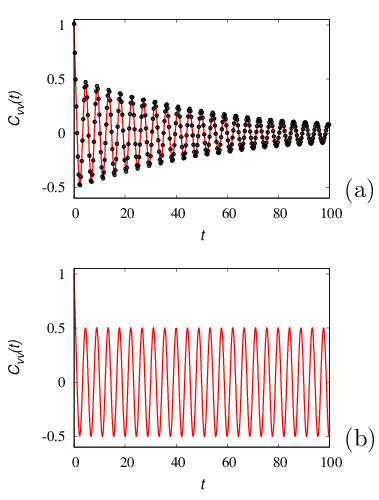}
\caption{Velocity autocorrelation function $C_{vv}(t)$ for the Plyukhin model. Panel (a) refers to
$\varepsilon=3.205$. Symbols ($\circ$) represent the results of stochastic simulations, the solid line
corresponds to the solution deriving from the fluctuation-dissipation relations of the first kind. Panel (b)
refers to $\varepsilon=\varepsilon^*$, i.e. to the critical condition for ergodicity breaking.}
\label{Fig6}
\end{figure}

As a second case study, consider the model analyzed by Plyukhin \cite{bao3}.
It corresponds to a memory kernel $h(t)$ of the form
\begin{equation}
h(t) = \delta(t) + \varepsilon \, e^{-\mu \, t} \, \sin ( \omega \, t)
\label{eqm_26}
\end{equation}
with $\mu=1/2$, $\omega=\sqrt{3}/2$. In \cite{bao3}, the critical value
of $\varepsilon$ giving rise to the ergodicity breaking is $\varepsilon^*=2 \, \sqrt{3}$.
In the present analysis we consider $\varepsilon$ as a parameter.
The stochastic realization of this system is expressed by
\begin{eqnarray}
\frac{d v(t)}{ dt} & = & - v(t) + \varepsilon \, z(t) + R_v(t) \nonumber \\
\frac{d r(t)}{d t} & = & - \mu \, r(t) + \omega \, z(t) + v(t) + R_r(t)
\label{eqm_27} \\
\frac{d z(t)}{ dt} & = & - \omega \, r(t) - \mu \, z(t) + R_z(t)
\nonumber
\end{eqnarray}
where $R_k(t)$, $\alpha=v,\,r,\,z$, are the stochastic forcings
\begin{equation}
R_k(t)= \sqrt{2} \, \left [  \sum_{h=1}^3 \beta_{k,h} \, \xi_h(t) \right ] \, , \qquad k=v,\,r,\,z
\label{eqm_28}
\end{equation}
and $\xi_h(t)$, $h=1,2,3$ are three independent white-noise processes (expressed
also in this case as distributional derivatives of independent Wiener processes).
Due to the presence of an impulsive contribution to friction, the stochastic forcing involves also
a contribution acting directly on the velocity variable $v(t)$.
The stability diagram  associated with the coefficient matrix of this system
vs the parameter $\varepsilon$ is represented in figure \ref{Fig4}. Specifically,
figure \ref{Fig4}  depicts the largest real part $\nu_{\rm max}^R$ of the eigenvalues
of the coefficient matrix ${\bf A}$,
\begin{equation}
{\bf A}=\left (
\begin{array}{ccc}
-1 & 0 & \varepsilon \\
1 & -\mu & \omega \\
0 & - \omega & -\mu 
\end{array}
\right )
\label{eqm_29}  
\end{equation}
and the global friction coefficient $\eta_\infty= 1+ \varepsilon \, \omega/(\mu^2+\omega^2)$.
As can be observed, the region of  dissipative stability coincides with the interval $(\varepsilon_{\rm min},\varepsilon^*)$,
where $\varepsilon_{\rm min}$ is the value of $\varepsilon$ at which $\eta_\infty=0$, and $\varepsilon^*$ is
the critical
point associated with the ergodicity breaking analyzed in \cite{bao3}. It remains to determine the region of stochastic realizability.
From eqs. (\ref{eqm_27})-(\ref{eqm_28}), considering for the white-noise processes the distributional
derivatives of independent Wiener processes, the associated Fokker-Planck equation  for the
density $p(v,r,z,t)$  reads
\begin{eqnarray}
\frac{\partial p}{\partial t} & = & - \frac{\partial \left [(- v + \varepsilon \, z) \, p \right ]}{\partial v}
- \frac{\partial \left [(- \mu \, r + \omega \, z+ v) \, p \right ]}{\partial r}
- \frac{\partial \left [(- \omega \, r - \mu \, z) \, p \right ]}{\partial z} \nonumber \\
& + & S_{v,v} \, \frac{\partial^2 p}{\partial v^2} +   2 S_{1,v} \, \frac{\partial^2 p}{\partial v \partial r}
+ 2 S_{2,v} \, \frac{\partial^2 p}{\partial v \partial z}+ S_{1,1} \, \frac{\partial^2 p}{\partial r^2}
+  2 S_{1,2} \, \frac{\partial^2 p}{\partial r \partial z} +  S_{2,2} \, \frac{\partial^2 p}{\partial z^2}
\label{eqm_30}
\end{eqnarray}
The system is stochastic realizable if there exists a positive definite symmetric matrix ${\bf S}$,
\begin{equation}
{\bf S}=
\left (
\begin{array}{ccc}
S_{v,v} & S_{1,v} & S_{2,v} \\
S_{1,v} & S_{1,1} & S_{1,2} \\
S_{2,v} & S_{1,2} & S_{2,2}
\end{array}
\right )
\label{eqm_31}
\end{equation}
such that eq. (\ref{eqm_30}) admits a stationary (equilibrium) density  $p^*(v,r,z)$ possessing the
following second-order moments $\langle v^2 \rangle_{\rm eq}=1$, $\langle v \, r \rangle_{\rm eq}= \langle
v \, z \rangle_{\rm eq}=0$, that ensure the correct behavior of the velocity autocorrelation
function consistently with the fluctuation-dissipation relation of the first kind.
It is convenient to introduce the following notation $m_{\alpha,\beta}(t)=\langle \alpha(t) \, \beta(t) \rangle$,
$\alpha,\beta=v,r,z$, for the second-order moments. Observe that in  this definition the
expected value $\langle \cdot \rangle$ refers to generic non-equilibrium conditions i.e. to
$p(v,r,z,t)$, and for this reason $m_{\alpha, \beta}(t)$ are functions of time. This should
not be confused with the expected value $\langle \alpha \, \beta \rangle_{\rm eq}$ that refers
to the equilibrium conditions (if they exist), i.e. to the invariant density $p^*(v,r,z)$ stationary solution of (\ref{eqm_30}).
It follows from eq. (\ref{eqm_30}) the system of moment equations
\begin{eqnarray}
\frac{d m_{v,v}}{d t} & = & - 2 \, m_{v,v}+ 2 \, \varepsilon \, m_{v,z} + 2 \, S_{v,v} \nonumber \\
\frac{d m_{v,r}}{d t} & = & - m_{v,r} + \varepsilon \, m_{r,z} - \mu \, m_{v,r} + \omega \, m_{v,z} + m_{v,v} + 2 \, S_{1,v}
\nonumber \\
\frac{d m_{v,z}}{d t} & = & - m_{v,z} + \varepsilon \, m_{z,z} - \omega \, m_{v,r} - \mu \, m_{v,z} + m_{v,v} + 2 \, S_{2,v}
\label{eqm_32} \\
\frac{d m_{r,r}}{d t} & = &  2  \left ( - \mu \, m_{r,r} +  \omega \, m_{r,z} + m_{v,r} \right ) + 2 \, S_{1,1} 
\nonumber \\
\frac{d m_{z,z}}{d t} & = & 2 \left ( - \omega \, m_{r,z} - \mu \, m_{z,z} \right ) + 2 \, S_{2,2} \nonumber \\
\frac{d m_{r,z}}{d t} & = & \left ( - \mu \, m_{r,z}+ \omega_{z,z} + m_{v,z} \right )
+ \left ( -\omega \, m_{r,r} -\mu \, m_{r,z} \right ) + 2 \, S_{1,2}
\nonumber 
\end{eqnarray}
Imposing the moment conditions, i.e. $m_{v,v}=1$, $m_{v,r}=m_{v,z}=0$ at equilibrium, we get from the moment equations
at steady state
\begin{equation}
S_{v,v}=1 \, , \quad m_{r,z} = - \frac{(1+2 \, S_{1,v})}{\varepsilon} \, , \quad
m_{z,z}= - \frac{2}{\varepsilon} \, S_{2,2}
\label{eqm_33}
\end{equation}
that imply $S_{1,v} \in {\mathbb R}$, and $S_{2,v}<0$ if $\varepsilon>0$. Moreover
\begin{equation}
m_{r,r}= \frac{1}{\mu} \left [ S_{1,1} - \frac{\omega}{\varepsilon} \, \left (1+ 2 \, S_{1,v} \right ) \right ]
\label{eqm_34}
\end{equation}
that implies that $S_{1,1} > S_{1,1}^*$, with $S_{1,1}^*=\omega \, \left (1+ 2 \, S_{1,v} \right ) /\varepsilon$.
As regards the remaining  entries of ${\bf S}$ we have,
\begin{eqnarray}
S_{2,2} & = & \omega \, m_{r,z}+ \mu \, m_{z,z} \nonumber \\
S_{1,2} & = & \frac{1}{2} \left [ 2 \, \mu \, m_{r,z} - \omega \, m_{z,z} + \omega \, m_{r,r} \right ]
\label{eqm_35}
\end{eqnarray}
The analysis of stochastic realizability can be developed in terms of an extensive search by
varying $S_{1,1} \in {\mathbb R}$, $S_{2,v} <0$ (since positive values of $\varepsilon$ are considered) and
$S_{1,1} > S_{1,1}^*$, determining the remaining entries of the ${\bf S}$ matrix from the expressions reported above, and
checking if the resulting ${\bf S}$ matrix is positive definite.
In this way, the region of stochastic realizability for the Plyukhin model can be determined.
Figure \ref{Fig5} depicts a projection of the region of stochastic realizability  with respect to $S_{2,2}$ as a function
of $\varepsilon$.  For any value of $\varepsilon$ (in the region of stochastic realizability), there exists an interval  $(S_{2,2}^-,S_{2,2}^+)$,
such that for $S_{2,2}$ falling within this interval there are conditions  (i.e. values of the other entries of ${\bf S}$) at which
the system is stochastically realizable. Observe that from simulations the system is stochastically realizable
up to a critical value $\varepsilon^{**}=3.449$, that is very close to  the Plyukhin threshold $\varepsilon^*=2 \, \sqrt{3}=3.464$.
The gap between $\varepsilon^{**}$ and $\varepsilon^{*}$ albeit small is net, but we cannot affirm with  certainty that this
gap  could not  be due to  tiny numerical effects in the analysis of the positive definiteness of ${\bf S}$.
This is indeed a minor detail in the present context.
To complete  the description of this case, figure \ref{Fig6} depicts the velocity autocorrelation function $C_{vv}(t)$ vs $t$
obtained from stochastic simulations of the Plyukhin dynamics for a value of $\varepsilon$ close to $\varepsilon^{**}$, compared
with the autocorrelation profile at the ergodicity-breaking threshold, i.e. at $\varepsilon^*$.
Given  a stochastic realization, i.e. an admissible positive semidefinite matrix derived from the moment equations
at steady state, the stochastic intensity matrix $\boldsymbol{\beta}$ is obtained, by assuming its symmetric nature,
from the equation $\boldsymbol{\beta}= {\bf S}^{1/2}$, that admits always a symmetric solution if ${\bf S}$ is
positive definite \cite{procopio}.

\section{Concluding remarks} 

The interpretation  of the phenomenon of ergodicity breaking
for  GLE  with well-behaved dissipative memory kernels within the broader dynamic theory
of GLE permits not only to appreciate better its dynamic origin, but also to set conditions
upon its physical manifestation.

By considering that in general the domain of stochastic realizability falls properly within
the  domain of dissipative stability, the  phenomenon of ergodicity breaking for well-behaved dissipative
kernels cannot be observed  from the solution of the associated stochastic differential equations,
and ultimately in physical systems. This statement has been proved for systems with kernel possessing two
real exponents, and this result can be generalized in the case of a larger number of decaying real modes  for which
$h(t)$ is for any $t>0$ greater than zero.
The Plyukhin model falls at the boundary of this analysis. But, apart from its mathematical and conceptual interest, its physical
validity should be further addressed and discussed, in order to ascertain whether it may represents a realistic
model for a physically realizable dynamics. Independently of this physical consistency check, its conceptual
relevance remains unaltered.

From the analysis developed in this article the importance of the
dynamic constraints in the analysis of GLE becomes evident: dissipative stability (and the
cases of ergodicity breaking analyzed by Bao, Plyukhin et al., represent a beautiful and conceptually
important example of it), and stochastic realizability.
A further remark on the latter property is important.

While dissipative stability is a property pertaining to the mean-field deterministic contribution
to the dynamics, and outside the region of dissipative stability the system does not admit an equilibrium behavior,
stochastic realizability is a more stringent condition (in most of the cases, as seen via the model
systems considered in this work) but, in the way it is defined, it is based on a physical assumption,
namely the validity of the Langevin condition eq. (\ref{eqm_4}). The Langevin condition  motivates the use
of white-noise processes in the representation of the thermal force, and is one-to-one with the expression for
the velocity autocorrelation function
stemming from the Kubo's fluctuation-dissipation relation of the first kind. In other words, the definition
of stochastic realizability adopted here is grounded on the validity of the Kubo theory.
The Langevin condition is not a fundamental principle of physics as the constant value of the velocity
of light in vacuo. It has represented an important condition to handle Brownian motion in fluids (gas and liquids)
in a given range of pressures and temperatures close to the ambient ones. But in principle it can be violated,
and its violation may lead to enlarge our understanding of fluctuational phenomena beyond the actual range.
In the eventuality of such an extension, the definition of stochastic realizability should be generalized accordingly.

\end{document}